\documentclass[fleqn,usenatbib]{mnras}

\usepackage{newtxtext,newtxmath}

\usepackage[T1]{fontenc}

\DeclareRobustCommand{\VAN}[3]{#2}
\let\VANthebibliography\thebibliography
\def\thebibliography{\DeclareRobustCommand{\VAN}[3]{##3}\VANthebibliography}

\usepackage{graphicx}	
\usepackage{amsmath}	
\usepackage{siunitx}
\usepackage{pifont}

\newcommand{\het}{He\,{\small II}}
\newcommand{\neth}{Ne\,{\small III}}
\newcommand{\ot}{O\,{\small II}}
\newcommand{\oth}{O\,{\small III}}

\title[High RO3 at $z=8.5$]{First Insights into the ISM at $z>8$ with JWST: Possible Physical Implications of a High [O\,{\small III}]\,$\mathbf{\lambda 4363}$/[O\,{\small III}]\,$\mathbf{\lambda 5007}$}

\author[H. Katz et al.] {Harley Katz$^{1}$\thanks{E-mail:
  \href{mailto:harley.katz@physics.ox.ac.uk}{harley.katz@physics.ox.ac.uk}}, Aayush Saxena$^2$, Alex J. Cameron$^1$, Stefano Carniani$^3$, Andrew J. Bunker$^1$, \newauthor Santiago Arribas$^4$,
  Rachana Bhatawdekar$^5$, Rebecca A. A. Bowler$^6$, Kristan N. K. Boyett$^{7,8}$, \newauthor Giovanni Cresci$^9$,  Emma Curtis-Lake$^{10}$,
  Francesco D'Eugenio$^{11,12}$, Nimisha Kumari$^{13}$, \newauthor Tobias J. Looser$^{11,12}$, Roberto Maiolino$^{11,12,2}$,  Hannah \"Ubler$^{11,12}$, Chris Willott$^{14}$, Joris Witstok$^{11,12}$
  \\
  $^1$Sub-department of Astrophysics, University of Oxford,
   Keble Road, Oxford OX1 3RH, United Kingdom \\
  $^2$Department of Physics and Astronomy, University College London, Gower Street, London WC1E 6BT, United Kingdom \\
  $^3$Scuola Normale Superiore, Universit\`a di Pisa, Piazza dei Cavalieri 7, I-56126 Pisa, Italy \\
  $^4$Centro de Astrobiolog\'ia (CAB), CSIC–INTA, Cra. de Ajalvir Km.~4, 28850- Torrej\'on de Ardoz, Madrid, Spain \\
  $^5$European Space Agency, ESA/ESTEC, Keplerlaan 1, 2201 AZ Noordwijk, NL \\
  $^6$Jodrell Bank Centre for Astrophysics, Department of Physics and Astronomy, School of Natural Sciences, The University of Manchester, Manchester, M13 9PL, UK \\
  $^7$School of Physics, University of Melbourne, Parkville 3010, VIC, Australia \\
  $^8$8ARC Centre of Excellence for All Sky Astrophysics in 3 Dimensions (ASTRO 3D), Australia \\
  $^9$INAF — Osservatorio Astrofisico di Arcetri, Largo E. Fermi 5, I-50125, Florence, Italy \\
  $^{10}$Centre for Astrophysics Research, Department of Physics, Astronomy and Mathematics, University of Hertfordshire, Hatfield, AL10 9AB, UK \\
  $^{11}$Kavli Institute for Cosmology, University of Cambridge, Madingley Road, Cambridge, CB3 0HA, UK\\
  $^{12}$Cavendish Laboratory - Astrophysics Group, University of Cambridge, 19 JJ Thompson Avenue, Cambridge, CB3 0HE, UK\\
  $^{13}$AURA for the European Space Agency, Space Telescope Science Institute, 3700 San Martin Drive, Baltimore, MD 21218, USA\\
  $^{14}$NRC Herzberg, 5071 West Saanich Rd, Victoria, BC V9E 2E7, Canada \\
  }

\date{Accepted XXX. Received YYY; in original form ZZZ}

\pubyear{2022}

\begin{document}
\label{firstpage}
\pagerange{\pageref{firstpage}--\pageref{lastpage}}
\maketitle

\begin{abstract}
We present a detailed analysis of the rest-frame optical emission line ratios for three spectroscopically confirmed galaxies at $z>7.5$. The galaxies were identified in the \emph{James Webb Space Telescope} (\emph{JWST}) Early Release Observations field SMACS J0723.3$-$7327. By quantitatively comparing Balmer and oxygen line ratios of these galaxies with various low-redshift ``analogue'' populations (e.g. Green Peas, Blueberries, etc.), we show that no single analogue population captures the diversity of line ratios of all three galaxies observed at $z>7.5$. We find that S06355 at $z=7.67$ and S10612 at $z=7.66$ are similar to local Green Peas and Blueberries. In contrast, S04590 at $z=8.50$ appears to be significantly different from the other two galaxies, most resembling extremely low-metallicity systems in the local Universe. Perhaps the most striking spectral feature in S04590 is the curiously high [O {\small III}]\,$\lambda4363$/[O {\small III}]\,$\lambda5007$ ratio (RO3) of $0.048$ (or $0.055$ when dust-corrected), implying either extremely high electron temperatures, $\sim3\times10^4$~K, or gas densities $>10^4\ {\rm cm^{-3}}$. Observed line ratios indicate that this galaxy is unlikely to host an AGN. Using photoionization modelling, we show that the inclusion of high-mass X-ray binaries or a high cosmic ray background in addition to a young, low-metallicity stellar population can provide the additional heating necessary to explain the observed high RO3 while remaining consistent with other observed line ratios. Our models represent a first step at accurately characterising the dominant sources of photoionization and heating at very high redshifts, demonstrating that non-thermal processes may become important as we probe deeper into the Epoch of Reionization.
\end{abstract}

\begin{keywords}
galaxies: high-redshift -- galaxies: evolution -- galaxies: ISM -- X-rays: binaries -- (ISM:) cosmic rays 
\end{keywords}



\section{Introduction}
The first stars, galaxies, and black holes to form at Cosmic Dawn directly impact all subsequent galaxy formation in the Universe. The stars in the first galaxies synthesize and distribute the first significant amounts of heavy elements, which modifies how gas can cool and form stars \citep[e.g.][]{Omukai2005}, while their ionizing photons heat and pressure smooth the intergalactic medium (IGM), impacting how galaxies can accrete gas from cosmic filaments \citep[e.g.][]{Okamoto2008,Katz2020}. Understanding the interstellar medium (ISM) within the first few generations of galaxies, and both the efficiency and properties of early star formation, is crucial for constraining the physics of the early Universe.

Rest-frame optical and UV emission line ratios provide robust diagnostics of the ISM conditions of galaxies, and the underlying photoionization or other energetic mechanisms that shape galaxy spectra.
Emission line ratios have been widely used at lower redshifts ($z \lesssim 3$) to distinguish photoionisation due to star-formation from active galactic nuclei (AGN) activity and other ionization sources (e.g. BPT diagnostics; \citealt{Baldwin1981, Kewley2001}) as well as to constrain the underlying physical conditions (e.g. density, metallicity, ionization parameter) within the ISM of galaxies \citep[e.g.][]{Kewley2019}. 
However, to date, the lack of sensitive space-based near-infrared spectroscopy has meant that detailed emission line studies of the ISM within the Epoch of Reionization ($z\gtrsim6$) have only been possible in the rest-frame far-infrared (most notably with ALMA), where few systems have been observed \citep[e.g.][]{Carniani2020}.
As a result, many studies have instead placed constraints on the ISM conditions of galaxies within the Epoch of Reionization using various samples of local galaxies deemed to be analogues of high-redshift galaxies (e.g. ``Green Peas''; \citealt{Cardamone2009, Yang2017a}, ``Blueberries''; \citealt{Yang2017b}, low-metallicity galaxies; \citealt{Izotov2019}, etc.).

The recent successful launch and commissioning of the \emph{James Webb Space Telescope} (\emph{JWST}; \citealt{Rigby2022}) is set to change this landscape.
In particular, the Near-Infrared Spectrograph (NIRSpec; \citealt{Jakobsen2022, Ferruit2022}) enables deep spectroscopy to be obtained across the full $\sim 1 - 5$ \micron \,wavelength range, allowing us to study rest-optical emission line ratios of $z\gtrsim6$ galaxies with unprecedented sensitivity.
Already, NIRSpec spectroscopic follow up of more than 40 new high-redshift galaxy candidates identified from NIRCam Early Release Observations (ERO; Program ID 2736, PI: K. Pontoppidan) of the SMACS J0723.3-7327 lensing cluster has revealed prominent detections of numerous rest-frame optical emission lines \citep{ERO2022}, enabling for the first time: (1) comparisons between the ISM conditions of ``high-redshift analogues'' and true $z\gtrsim6$ galaxies; and (2) direct constraints on the ISM conditions of $z\gtrsim6$ galaxies from rest-optical emission lines.

To constrain the formation and evolution of some of the earliest structures in the Universe, it is vital to compare the observed spectroscopic properties with predictions from models. This includes both photoionization models of the ISM \citep[e.g.][]{Gutkin2016,Nakajima2022}, or numerical simulations that attempt to model the relevant physics from first principles \citep[e.g.][]{Katz2019}. Such comparisons will allow us to accurately infer the dominant sources of ionising photons and the important cooling and heating processes in high-redshift galaxies. To this end, in this work, we investigate the rest-frame optical emission lines in three galaxies at $z>7.5$ in the SMACS0723 field and compare their spectroscopic properties with various low-redshift galaxy populations and infer the physics that is governing the ISM in the early Universe.

The layout of this paper is as follows: we discuss the \emph{JWST} NIRSpec ERO data used in this study, along with reduction steps, properties of the three $z>7.5$ galaxies and emission line fitting methodology in \S\ref{sec:data}. We discuss insights obtained from emission line ratios and compare the observations of galaxies at $z>7.5$ with those in the local Universe in \S\ref{sec:lines}. In \S\ref{sec:curious}, we explore the physics that can generate the curiously high [\oth]\,$\lambda4363$/[\oth]\,$\lambda5007$ ratio observed at $z=8.5$ using a suite of photoionization models. Our discussion and conclusions are presented in \S\ref{sec:conclusions}.

\section{Data and emission line fitting}
\label{sec:data}
\begin{figure*} 
\centering
    \includegraphics[width=\textwidth]{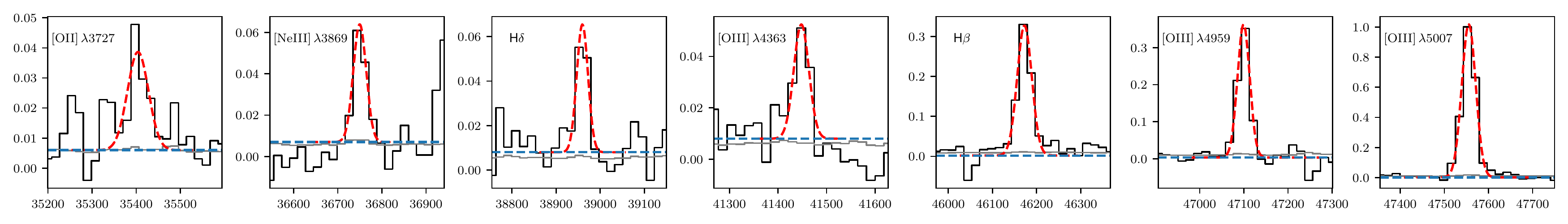}
    \includegraphics[width=0.93\textwidth]{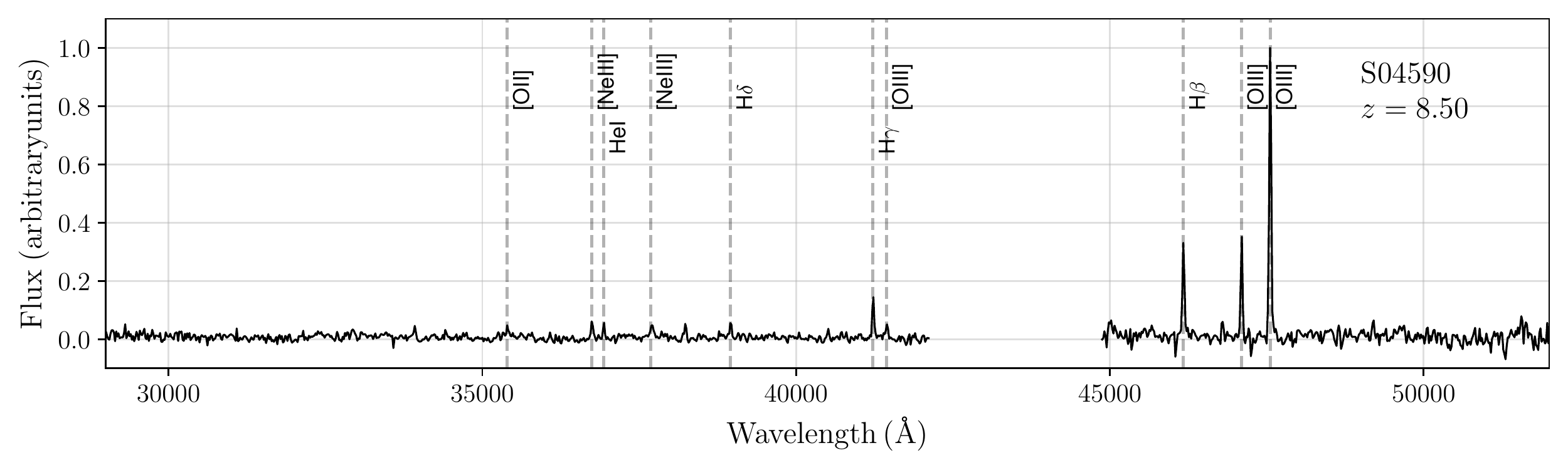}

    \vspace{-3pt}
    \includegraphics[width=\textwidth]{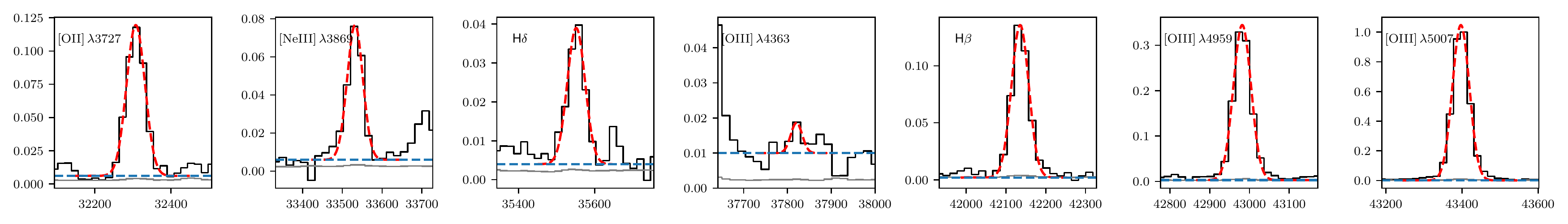}
    \includegraphics[width=0.93\textwidth]{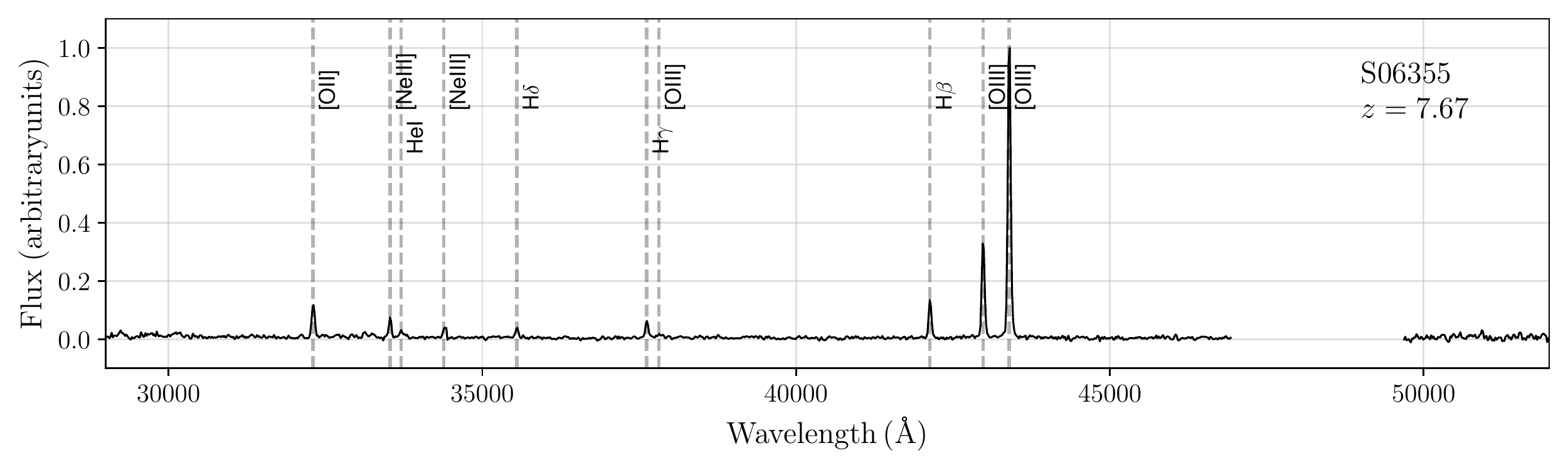}
    
    \vspace{-3pt}
    \includegraphics[width=\textwidth]{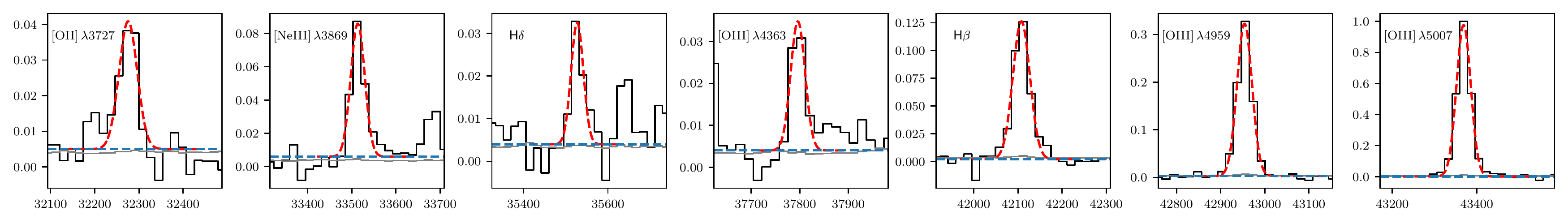}
    \includegraphics[width=0.93\textwidth]{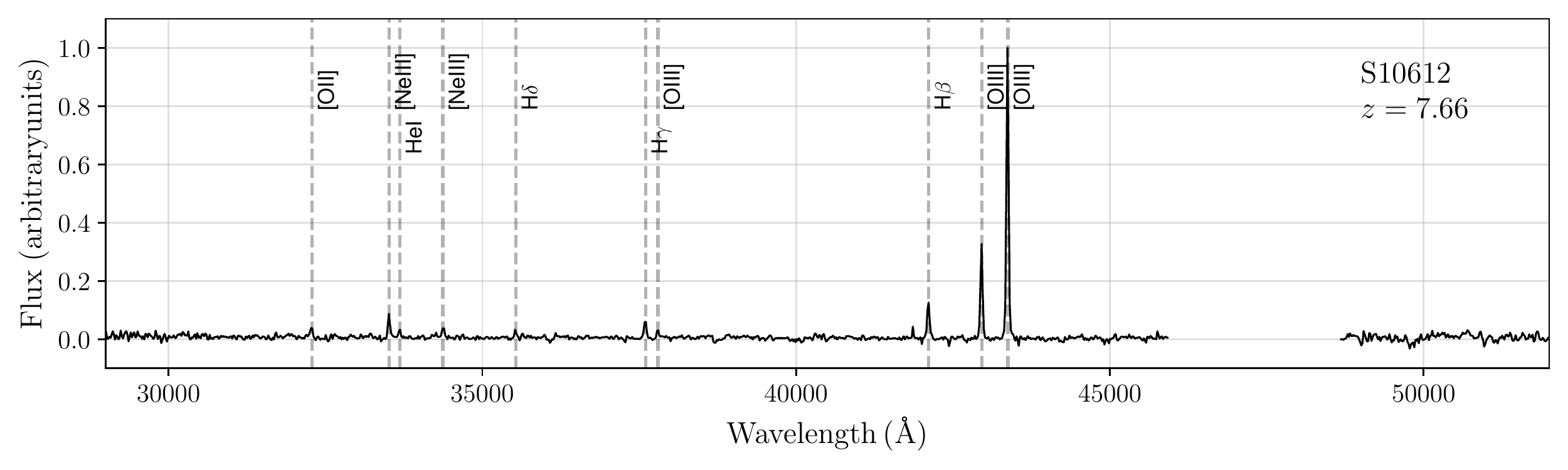}
    
    \vspace{-10pt}
    \caption{Observed frame NIRSpec spectra of S04590 at $z=8.498$ (top), S06355 at $z=7.666$ (middle) and S10612 at $z=7.663$ (bottom). The spectra have been normalised such that the maximum flux value is at 1. The locations of emission lines have been marked for each source and the best-fitting Gaussians (red) are shown above the full 1D spectra for each object with the same axis labels as the 1D spectra, where we also show the base of the Gaussian fit (blue) inferred from the local continuum around the line, broadly consistent with the global noise in the 1D spectra (gray). The missing data is due to the gap between the two NIRSpec detector arrays.}
    \label{fig:spectra}
\end{figure*}


We use data from \emph{JWST} NIRSpec Micro-Shutter Assembly (MSA) observations of the SMACSJ0723.3-7327 galaxy cluster taken on 2022 Jun 30, as part of the public ERO data release (Program ID 2736, PI: K. Pontoppidan, \citealt{ERO2022}). The $0.2^{\prime\prime} \times 0.46^{\prime\prime}$ MSA shutters are configurable and can be used to acquire spectra of hundreds of sources in a single field-of-view \citep{Ferruit2022}.
The spectra were taken in two filter/disperser combinations: F170LP/G235M and F290LP/G395M which combine to span from $1.66-5.10$\,$\mu$m with a resolving power of $R\sim1000$.
The MSA configuration involved placing slitlets comprised of a column of at least three consecutive open shutters in a single column on each target. Two 8,753~second observations were obtained for this field (Obs~007 and Obs~008 hereafter), each of which involved a three-point nodding pattern to enable local background subtraction, for a total of $\sim$17,507 seconds.

Our analysis in this paper is performed on the publicly available\footnote{\url{https://doi.org/10.5281/zenodo.6940561}} extracted 1D spectra presented in \citet{Curti2022}.
These resulted from processing Stage 1 outputs from the JWST data calibration pipeline with the GTO pipeline (NIRSpec GTO collaboration, in prep.).
These data were reduced assuming a point-like source profile, although from NIRCam imaging we see that the sources are marginally resolved. While this assumption may introduce some uncertainty on the absolute flux calibration, the analysis presented in this paper is based only on emission line flux ratios. We estimate that the assumed source profile affects these flux ratios at the $<10\%$ level.
Background subtraction was performed locally by subtracting the background flux as measured by neighbouring ``empty'' shutters.
The reader is directed to \citet{Curti2022} for specific details about the calibration of these data.
Finally, we note that, although we used the same 1D extracted spectra, the emission line fitting presented here was done independently of that in \citet{Curti2022}.

This study focuses on three objects confirmed to lie at $z>7.5$: S04590 at $z=8.50$ (RA: $110.89533$, Dec: $-73.44916$), S06355 at $z=7.67$ (RA: $110.84452$, Dec: $-73.43508$), and S10612 at $z=7.66$ (RA: $110.89533$, Dec: $-73.43454$). These objects are all gravitationally lensed by the foreground cluster; however, crucially for this study, the achromatic nature of gravitational lensing means that the emission line ratios are magnification-independent, and therefore, the results presented here are not affected by magnification at all. Extracted 1D spectra of all three objects along with the identified emission lines are shown in Figure \ref{fig:spectra}.

We run emission line fitting for the following rest-optical emission lines, identified as being detected in the spectra: [\ot] $\lambda\lambda$3726, 3729, [\neth] $\lambda$3869, H$\delta$, H$\gamma$, [\oth] $\lambda$4363, H$\beta$, [\oth] $\lambda$4959 and [\oth] $\lambda$5007. We note that all of the emission lines used in this study are measured in the F290LP/G395M filter/disperser setting, so there is no uncertainty introduced by comparing the relative calibration of these two settings.
We find the emission line profiles can be well fit by a single Gaussian function and use the best fitting parameters to infer line fluxes and ratios. 

To obtain the best-fitting Gaussian, we model the continuum (where detected) or noise level around each emission line as a linear fit and measure its median and standard deviation. Best-fitting Gaussians (red) for the main emission lines investigated in this study are shown in Figure \ref{fig:spectra}. Although we do not force the width of the Gaussians to be consistent across all emission lines fitted, we find good agreement in the measured full-width at half maxima (FWHM) of the best-fitting Gaussians ($\approx200-250$\,km\,s$^{-1}$, consistent with being unresolved at $R\sim1000$). We find that our ``local'' estimates of the continuum/noise that we use to fit Gaussians (blue dashed line) are broadly consistent with the noise estimates in the spectra returned by the data reduction pipeline (gray solid line) in line with expectations from weak to no continuum detection. A notable exception is the [\oth]\,$\lambda4363$ in S06355 (middle panel in Figure \ref{fig:spectra}), where two noise peaks appear to straddle the [\oth]\,$\lambda4363$ peak.

We find secure detections of H$\beta$, [\oth]\,$\lambda4959$, and [\oth]\,$\lambda5007$ across all three galaxies. The [\ot]\,$\lambda \lambda\,3726,3729$ doublet is unresolved, and is weak with respect to [\oth]\,$\lambda 5007$, as evidenced by the relatively high [\oth]/[\ot] ratios (see Table \ref{tab:ratios}) for all three galaxies. Interestingly, the [\oth]\,$\lambda4363$ auroral line is securely detected in S04590 and S10612 with ${\rm S/N}\approx5.3$ and $\approx6.3$, respectively, with a lower ${\rm S/N}\approx2.8$ detection in S06355. We also securely detect H$\delta$ and H$\gamma$ in the galaxies, in addition to the [\neth]\,$\lambda 3869$ and $\lambda3967$ lines, although the [\neth]\,$\lambda3967$ line is blended with H$\epsilon$. 

We then correct the emission line fluxes for the effects of dust extinction based on the observed Balmer decrement H$\gamma$/H$\beta$, applying the appropriate dust correction following the Small Magellanic Cloud extinction curve from \citet{Gordon2003}, assuming an intrinsic value of $0.468$ corresponding to $n_{\rm e} = 100$\,cm$^{-3}$ and $T=10^4$\,K \citep[e.g.][]{Osterbrock2006}. For S04590 we find that dust correction of with $A_{\rm V} = 0.64 \pm 0.20$ is required to achieve a Balmer decrement of $0.468$\footnote{As we discuss below, the electron temperature measured for S04590 is $>10^4$~K. If we instead apply a dust correction assuming a Balmer decrement value at $n_{\rm e} = 100$\,cm$^{-3}$ and $T=3\times10^4$\,K, the dust-corrected value of RO3 increases to 0.062 for S04590.}, in line (albeit slightly lower) with what was reported by \citet{Curti2022}. This dust extinction value is consistent with the low-metallicity, low-stellar mass galaxies at high-redshift in the SPHINX$^{20}$ simulation \citep{Katz2022a}. We note that even without a dust correction, the RO3 value for this particular galaxy is much higher than is typically observed. We find that the other two galaxies require no dust correction, i.e. $A_V < 0.1$. 

Despite employing slightly different emission line measurement techniques, the emission line ratios we measure are in agreement with those reported by \citet{Curti2022}. The emission line ratios and their uncertainties employed in this study are given in Table \ref{tab:ratios}.

\begin{table*}
    \centering
    \caption{Observed rest-frame optical emission line ratios corrected for dust extinction.}
    \begin{tabular}{lccc}
\hline
Ratio & S04590 & S06355 & S10612 \\
\hline

[\oth]\,$\lambda4363$/[\oth]\,$\lambda 5007$ (RO3) & $0.055 \pm 0.010$ & $0.010 \pm 0.009$ & $0.033 \pm 0.005$ \\


[\oth]\,$\lambda 5007$ / [\ot]\,$\lambda\lambda 3726, 3729$ (O32) & $15.781 \pm 3.040$ & $8.737 \pm 0.295$ & $20.733 \pm 2.818$ \\


[\oth]\,$\lambda5007$/H$\beta$ & $3.239 \pm 0.260$ & $8.603 \pm 0.220$ & $6.820 \pm 0.420$ \\


[\neth]\,$\lambda 3869$/[\ot]\,$\lambda\lambda 3726, 3729$ & $1.392 \pm 0.450$ & $0.518 \pm 0.035$ & $1.654 \pm 0.247$ \\


[\ot]\,$\lambda\lambda 3726, 3729$/H$\delta$ & $1.020 \pm 0.264$ & $3.101 \pm 0.303$ & $2.075 \pm 0.572$ \\


[\neth]\,$\lambda 3869$/H$\delta$ & $1.421 \pm 0.445$ & $1.607 \pm 0.179$ & $3.433 \pm 0.858$ \\

H$\gamma$/H$\beta$ (observed, not dust corrected) & $0.408\pm0.026$ & $0.467\pm0.016$ & $0.468\pm0.038$ \\

\hline
    \end{tabular}
    \label{tab:ratios}
\end{table*}

\begin{figure*}
\centerline{
\includegraphics[scale=1,trim={0 0.0cm 0cm 1.2cm},clip]{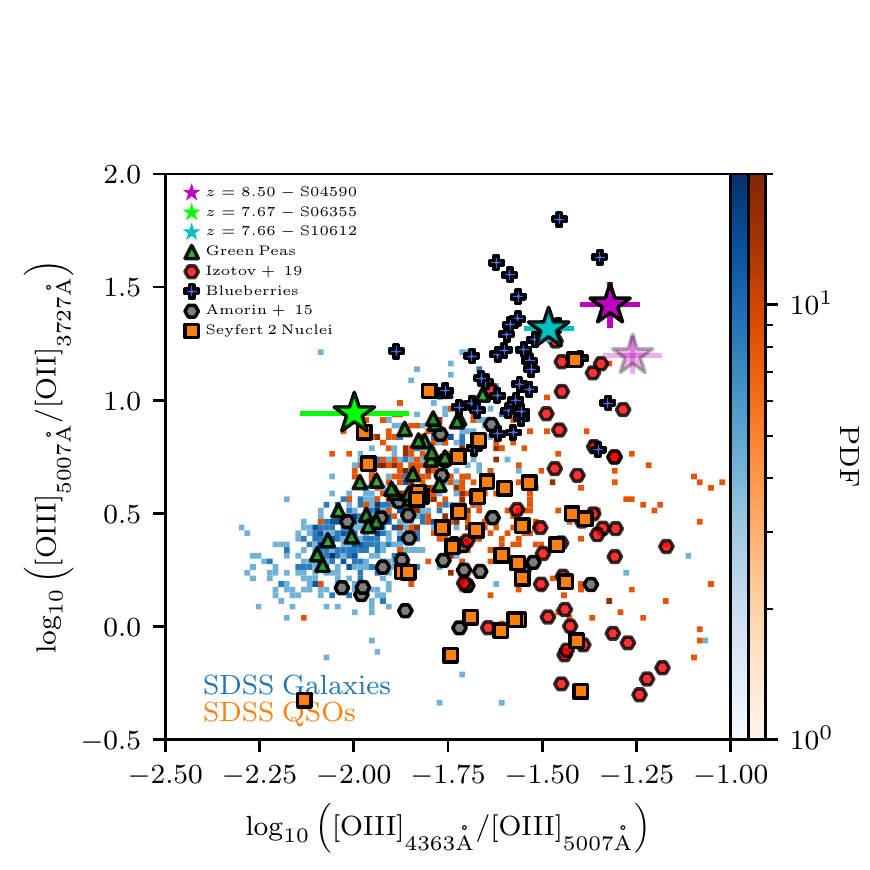}
\includegraphics[scale=1,trim={0 0.0cm 0cm 1.2cm},clip]{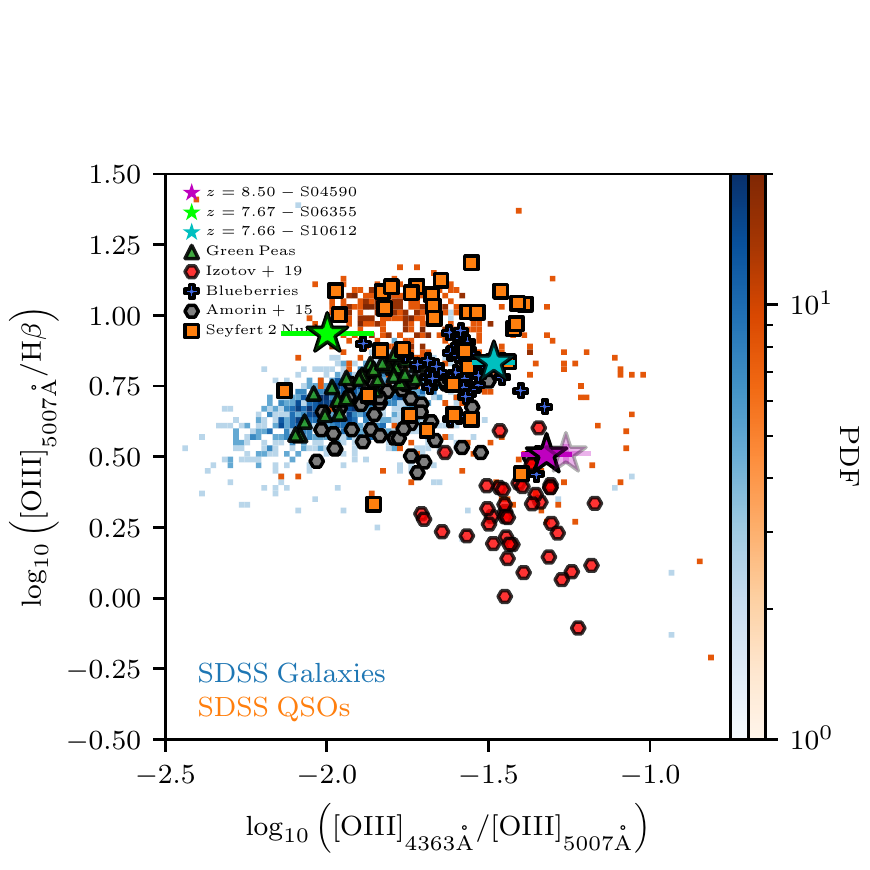}
}
\centerline{
\includegraphics[scale=1,trim={0 0.0cm 0cm 1.2cm},clip]{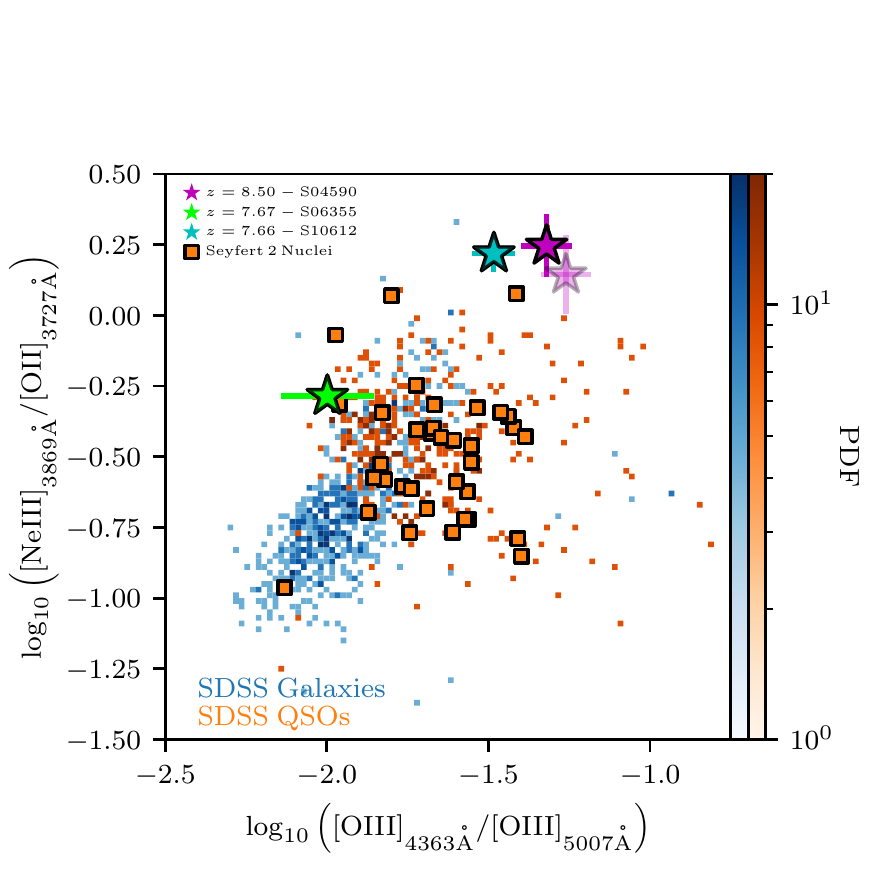}
\includegraphics[scale=1,trim={0 0.0cm 0cm 1.2cm},clip]{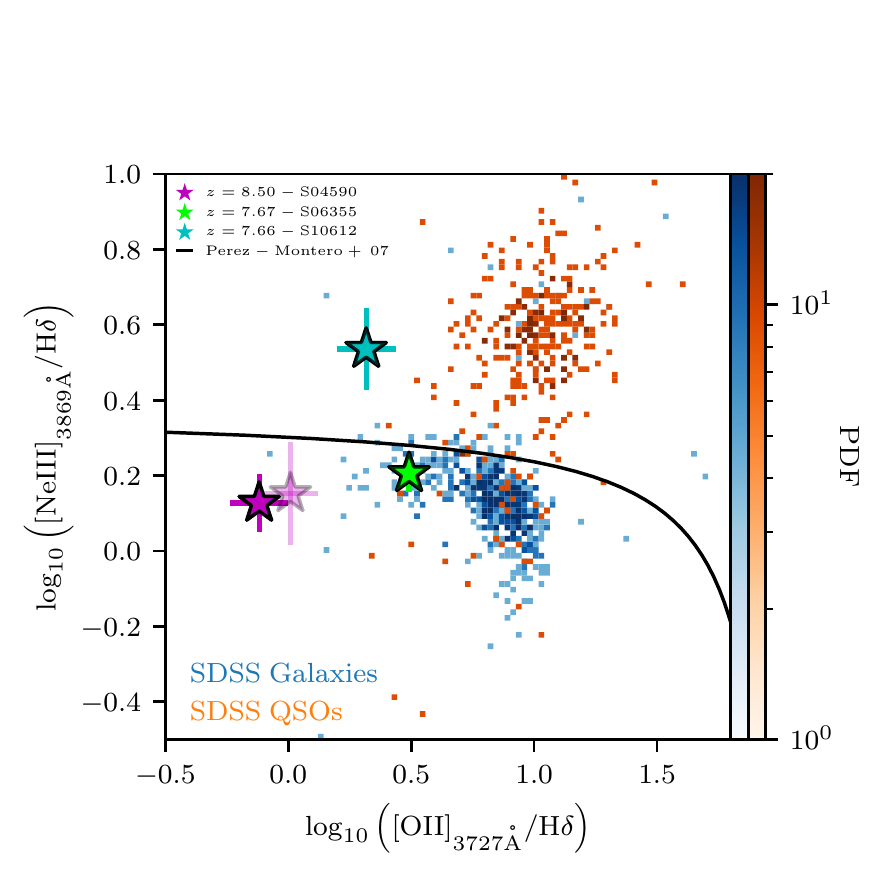}
}
\caption{({\bf Top Left}) RO3 versus O32. The three high redshift galaxies are shown as coloured stars with error bars representing the 1$\sigma$ uncertainty on the flux ratios. Translucent data points have been corrected for dust. For comparison, we show 2D probability distribution functions (PDFs) of SDSS galaxies (blue) and SDSS QSOs (orange) for objects that have at least an 8$\sigma$ detection of the [\oth]\,$\lambda 4363$ line. Low-redshift analogues of high-redshift galaxies including green peas \protect\citep{Yang2017a}, blueberries \protect\citep{Yang2017b}, extremely low metallicity galaxies \protect\citep{Izotov2019}, and extreme emission line galaxies \protect\citep{Amorin2015} are shown at green triangles, blue crosses, red hexagons, and grey hexagons, respectively. Orange squares are a compilation of Seyfert 2 nuclei \protect\citep{Armah2021}. ({\bf Top Right}) RO3 versus [\oth]\,$\lambda 5007$/H$\beta$. ({\bf Bottom Left}) RO3~versus [\neth]\,$\lambda 3869$/[\ot]\,$\lambda\lambda 3726, 3729$. ({\bf Bottom Right}) [\ot]\,$\lambda\lambda 3726, 3729$/H$\delta$ versus [\neth]\,$\lambda 3869$/H$\delta$. The thick black line is the AGN/star-forming galaxy demarcation as described in \protect\cite{PZ2007}. Unless otherwise specified, the symbols in all panels are the same.}
\label{fig1}
\end{figure*}

\section{Insights from rest-frame optical emission lines}
\label{sec:lines}
We start our analysis of the observed emission line ratios of the three $z>7.5$ galaxies by comparing them with various populations of lower redshift galaxies. Our goal is to address the question: how similar are reionization-epoch galaxies to low-redshift galaxies and to local analogues of high-redshift galaxies? 

In Figure~\ref{fig1} we show various strong-line diagnostics: [\oth]\,$\lambda4363$/[\oth]\,$\lambda 5007$ (RO3) vs. [\oth]\,$\lambda 5007$/[\ot]\,$\lambda\lambda 3726, 3729$ (O32; top left); RO3 vs. [\neth]\,$\lambda 3869$/[\ot]\,$\lambda\lambda 3726, 3729$ (bottom left); RO3 vs. [\oth]\,$\lambda5007$/H$\beta$ (top right); and [\ot]\,$\lambda\lambda 3726, 3729$/H$\delta$ vs. [\neth]\,$\lambda 3869$/H$\delta$ (bottom right). In all panels, we show emission line ratio measurements from SDSS\footnote{Emission line ratios have been adopted from the MPA-JHU DR7 catalog: \url{https://www.mpa-garching.mpg.de/SDSS/DR7/}} for galaxies and quasars, split based on the \cite{Kauffmann2003} BPT criteria, selecting only those sources that have $\ge 8\sigma$ detection of [\oth]\,$\lambda4363$. For additional comparison, where data is available, we show the locations of low-redshift Seyfert~2 nuclei \citep{Armah2021}, low redshift analogues in the form of Green Peas \citep{Yang2017a}, Blueberries \citep{Yang2017b}, and extremely low-metallicity galaxies ($12+\log({\rm O/H})\lesssim7.5$; \citealt{Izotov2019}), as well as extreme emission-line galaxies \citep{Amorin2015}.

Interestingly, we find that there is no single low-redshift reference sample that can perfectly explain the properties of all three $z>7.5$ galaxies (when selecting only those that exhibit [\oth]~$\lambda$4363). Within error bars, S06355 at $z=7.67$ (lime green star) is qualitatively consistent with Green Peas, Blueberries, and the edge of the distribution of SDSS galaxies. 

Based on oxygen emission lines and H$\beta$, S10612 at $z=7.66$ (cyan star) is consistent with Blueberries, but seems to have higher O32 and RO3 compared to the Green Peas, and significantly higher [\oth]~$\lambda$5007/H$\beta$ compared to the extremely low-metallicity galaxies from \cite{Izotov2019}. This galaxy also shows exceptionally strong [\neth]~$\lambda$3869, falling above the AGN demarcation line from \cite{PZ2007} when comparing [\neth]~$\lambda$3869/H$\delta$ vs [\ot]~$\lambda\lambda$3726, 3729/H$\delta$. We show below that such emission line ratios can also be driven by strong shocks, so this galaxy is not conclusively an AGN, nor has a signature of higher excitation energy emission lines such as \het~$\lambda$4686. 

Similar to S06355, S04590 at $z=8.5$ (magenta star) shows very high O32, RO3, and [\neth]~$\lambda$3869/[\ot]~$\lambda\lambda$3726, 3729. This galaxy resides at the very edges of the distributions of Blueberries and extremely low-metallicity galaxies. The dust-corrected RO3 value of S04590, at 0.059, is among the highest currently known in the local Universe. Unlike S06355, S04590 is characterised by very high Balmer emission compared to the emission from metal lines, with [\oth]~$\lambda$5007/H$\beta$ lying below the typical values in SDSS. Furthermore, both [\neth]~$\lambda$3869/H$\delta$ and [\ot]~$\lambda\lambda$3726, 3729/H$\delta$ are low compared to the typical SDSS galaxy, as can be seen in the bottom right panel of Figure~\ref{fig1}.

\begin{table}
    \centering
    \caption{Probability of being consistent with the reference sample of galaxies as computed by a trained isolation forest.}
    \begin{tabular}{lccc}
\hline
Galaxy Sample & S04590 & S06355 & S10612 \\
\hline
SDSS & 0.00 & 0.53 & 0.06 \\
Green Peas & 0.00 & 0.54 & 0.46 \\
Blueberries & 0.01 & 0.91 & 1.00 \\
Izotov+ 2019 & 0.96 & 0.00 & 0.06 \\
Amorin+ 2015 & 0.00 & 0.00 & 0.00 \\
\hline
    \end{tabular}
    \label{tab:iso}
\end{table}

To quantify the similarity between these three high-redshift galaxies and each low-redshift reference population, we train an Isolation Forest \citep{Liu2008} on the three emission line ratios of O32, RO3, and [\oth]~$\lambda$5007/H$\beta$. We compute the probability that each $z>7.5$ galaxy is an ``inlier'' for each reference galaxy population by randomly sampling the three flux ratios 10,000 times across the error distribution of the reference populations and using the Isolation Forest to classify\footnote{This classification requires a choice of ``contamination''. For all reference samples, we set the contamination to 2\%. The results are qualitatively similar if we choose 1\%.} the galaxy. 

In Table~\ref{tab:iso} we report the fraction of these 10,000 samples that are classified as inliers for each reference population. Consistent with our qualitative analysis above, we see that S06355 at $z=7.67$ is marginally consistent with SDSS galaxies, Green Peas, and consistent with the spread in properties exhibited by Blueberries, while S10612 at $z=7.66$ is only consistent with Green Peas and Blueberries. In contrast, S04590 at $z=8.50$ is most similar to the extremely low metallicity galaxies of \cite{Izotov2019} and sits at the very edge of the distribution of Blueberries. This quantitatively demonstrates that these three high-redshift galaxies, despite lying at similar redshifts, are diverse in nature, and multiple different low-redshift analogue samples are needed to realistically encapsulate their properties (cf. \citealt{Schaerer2022,Rhoads2022}).

Since all three galaxies have high O32 and RO3 compared to SDSS galaxies, the ISM in these high-redshift systems likely have high ionization parameters, lower metallicities, and/or higher H{\small II} region electron temperatures and densities compared to that in the local Universe (see also \citealt{Trump2022}). However, with this initial analysis, using the dust-corrected values, two key features stand out: the very high RO3 in S04590 and the very high [\neth]~$\lambda$3869/H$\delta$ and O32 in S10612. The latter can likely be explained by a high ionization parameter and low metallicity; however, the former requires more investigation, which we discuss further below.

\section{Curiously high [\oth]\,$\mathbf{\lambda4363}$/[\oth]\,$\mathbf{\lambda5007}$ ratio at Redshift 8.5}
\label{sec:curious}
Qualitatively, the spectral feature that is immediately interesting is the strength of the [\oth]\,$\lambda 4363$ line with respect to [\oth]\,$\lambda 5007$ in S04590 at $z=8.5$. For galaxies in SDSS with measured [\oth]\,$\lambda 4363$, we find $\log_{10}({\rm RO3})=-1.91\pm0.3$ (i.e. $0.012$). The value is larger for low-metallicity galaxies ($-1.41$ or $0.039$; \citealt{Izotov2019}), however, S04590 has a $\log_{10}$(RO3) of $-1.32$ ($-1.26$ when corrected for dust), which is a $2\sigma$ outlier compared to the SDSS population with detected [\oth]\,$\lambda 4363$.

\begin{figure}
\centerline{
\includegraphics[scale=1,trim={0 0.0cm 0cm 1cm},clip]{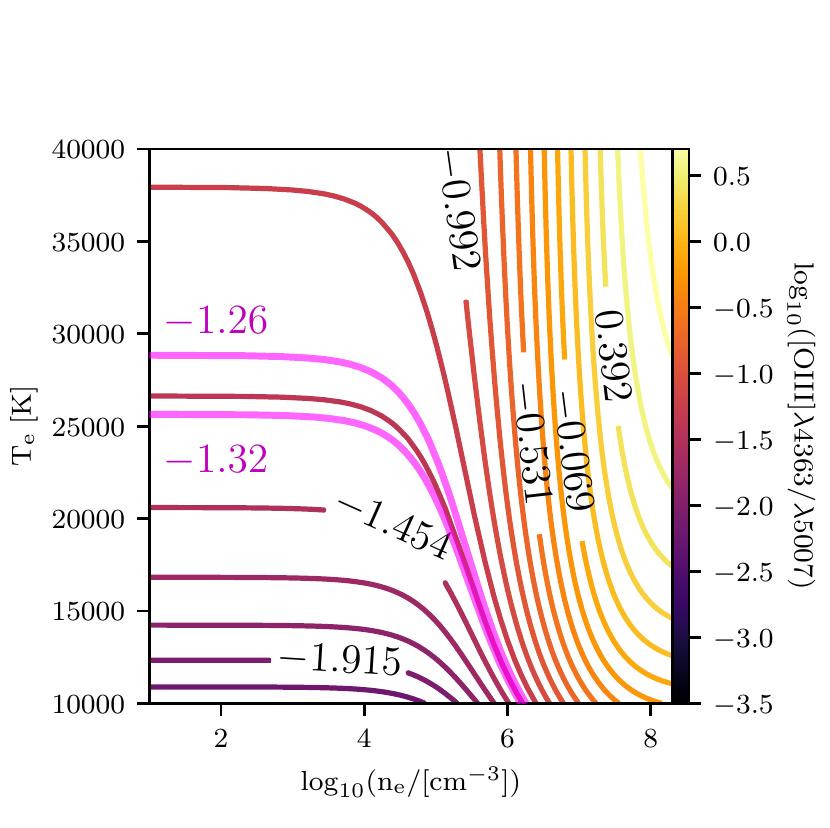}
}
\caption{Contours of constant RO3 as a function of electron density and temperature. The colour of the line represents RO3 as given in the colour bar. The two thick magenta lines represent the observed (lower) and dust-corrected (upper) contours that best match S04590.}
\label{fig2}
\end{figure}

Figure~\ref{fig2} shows contours of constant RO3 as a function of electron temperature and density, demonstrating the physical conditions required to give rise to this high RO3 value. High RO3 values can either be driven by high electron temperatures ($T_{\rm e}$) or high electron densities ($n_{\rm e}$). The change in level populations with temperature drives the behaviour of the ratio with temperature, while the dependence on density is caused by the fact that the critical density of the $\lambda 5007$ line is lower than that of the $\lambda 4363$ line \citep[e.g.][]{Osterbrock2006}. 

In order to obtain the dust-corrected value of RO3 ($\sim 0.055$) as is seen in S04590, one either requires temperatures $\sim3\times10^4$~K, or densities $>10^4\ {\rm cm^{-3}}$. Such a temperature is higher than what is typically seen in galactic H{\small II} regions \citep[e.g.][]{Deharveng2000} where $T_{\rm e}\sim7000-14000$~K and $n_{\rm e}\sim100\ {\rm cm^{-3}}$ \citep{Osterbrock2006}, and those computed from photoionization models for typical stellar SED choices \citep[e.g.][]{Xiao2018}, while such electron densities are an order of magnitude greater than those seen in star-forming galaxies at $z\sim2$, where H{\small II} region densities can be directly inferred \citep[e.g.][]{Sanders2016}. To demonstrate this further, in the top row of Figure~\ref{cr_xrb}, we show the results of a set of {\small CLOUDY} models \citep{Ferland2017} where we have varied the stellar age, ionization parameter, and metallicity of the gas and stellar population that is illuminating the cloud (see Appendix~\ref{cloudy} for details on the models). Only the models with the lowest metallicities ($10^{-2}Z_{\odot}$) and the highest ionization parameters ($10^{-1.5}$) begin to approach the values of RO3 and O32 seen in S04590; however, these models cannot simultaneously reproduce the strength of [\oth]~$\lambda$5007/H$\beta$ due to the low metallicity. Hence, under relatively standard assumptions (see Appendix~\ref{bpe} for an extended set of models) on H{\small II} region properties, it is difficult to reconcile what is driving the emission line ratios observed in S04590.

Before investigating the physical conditions required to drive such a high RO3, we reiterate the fact that \emph{JWST}/NIRSpec is a new instrument and the observations underpinning this work are among the first taken at such high redshifts. We note that in an independent analysis of these spectra, \citet{Curti2022} reported a slightly lower value of ${\rm RO3}=0.045\pm0.007$ for S04590 (before dust correction). While this value is lower than what we find here (0.048), we emphasise that the models described above are still unable to simultaneously reproduce those RO3 and O32 ratios reported in \citet{Curti2022}. If further observations continue to find such high RO3 ratios in high-redshift galaxies, there must be some additional physics beyond star formation that is responsible. We consider four possible solutions below:
\begin{enumerate}
    \item The presence of an AGN
    \item Strong shocks
    \item X-ray heating
    \item Cosmic ray heating
\end{enumerate}

\subsection{The presence of an AGN?}
The observation of high RO3 is reminiscent of the temperature problem in low-redshift Seyfert galaxies \citep[e.g.][]{Koski1976,Osterbrock1978,Ferland1983,Dopita1995,Nagao2001,Baskin2005,Binette2022}. The narrow line region (NLR) around an AGN can reach the densities and temperatures that are perhaps required to explain the high RO3 in S04590 (e.g. as was shown in the photoionization models of \citealt{Baskin2005}). The line widths of S04590 in the spectrum remain spectrally unresolved at $R\sim1000$, resulting in an upper limit of $<250$\,km\,s$^{-1}$, placing them on the low-end of what one might expect for an NLR. Nevertheless, we first consider an AGN as a potential solution for the high RO3 ratio.

Not only is S04590 inconsistent with SDSS QSOs and low-redshift Seyfert~2 nuclei on all diagnostics considered, it falls well below the AGN-galaxy demarcation proposed by \citep{PZ2007} on the [\ot]~$\lambda\lambda$3726, 3729/H$\delta$ versus [\neth]~$\lambda$3869/H$\delta$ diagnostic. Without a measurement of [N{\small II}]~$\lambda$6583, H$\alpha$, we cannot distinguish between star formation and AGN activity using the BPT diagram. Furthermore high excitation lines such as He{\small II} lines\footnote{The optical \het~$\lambda$4686 line falls in a band gap.}, or any of the higher ionization state Fe lines (these are likely too weak to be detected) are absent from the spectrum, making it hard to conclusively determine whether this galaxy hosts an AGN. However, none of the available evidence suggests that it does host an AGN. For these reasons, we consider alternative solutions to explain the spectrum of this object.

\subsection{Strong shocks}
Fast, radiative shocks can also increase the temperature of the gas well above $10^4$~K \citep{Allen2008, Dopita2017}, potentially in the regime needed to explain the high RO3 in S04590. To determine whether such shocks can plausibly reproduce the emission line ratios we see at $z>7.5$, we compare line ratios from the {\small MAPPINGS} models of \cite{Allen2008} to the three high-redshift galaxies in Figure~\ref{shocks} across the four diagnostics presented above. 

As before, we find that S04590 remains a clear outlier even when compared with these shock models. Therefore, we conclude that shocks alone are unlikely to explain the emission line ratios seen in this particular galaxy. We highlight that shocks can explain the [\neth]~$\lambda$3869/H$\delta$ and [\ot]~$\lambda\lambda$3726, 3729/H$\delta$ ratios seen in S10612, suggesting that, although this galaxy lies above the \citet{PZ2007} AGN demarcation line in this diagnostic plot (lower-right panel of Figure~\ref{fig1}), the line ratios can be explained without invoking the presence of an AGN.

\begin{figure*}
\centerline{
\includegraphics[scale=1,trim={0 0.0cm 0cm 0cm},clip]{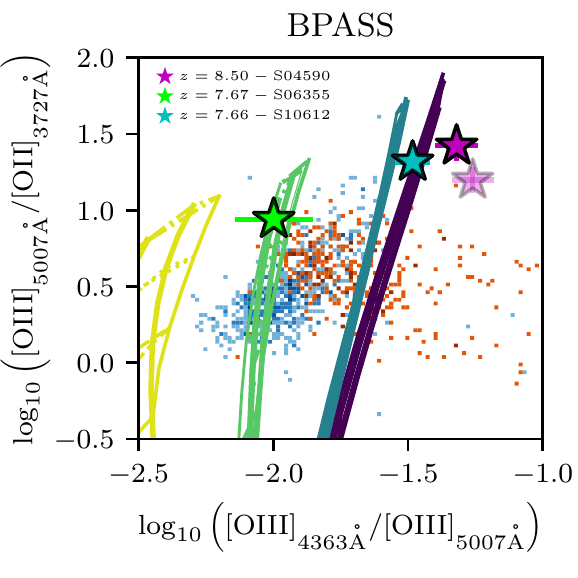}
\includegraphics[scale=1,trim={0 0.0cm 0cm 0cm},clip]{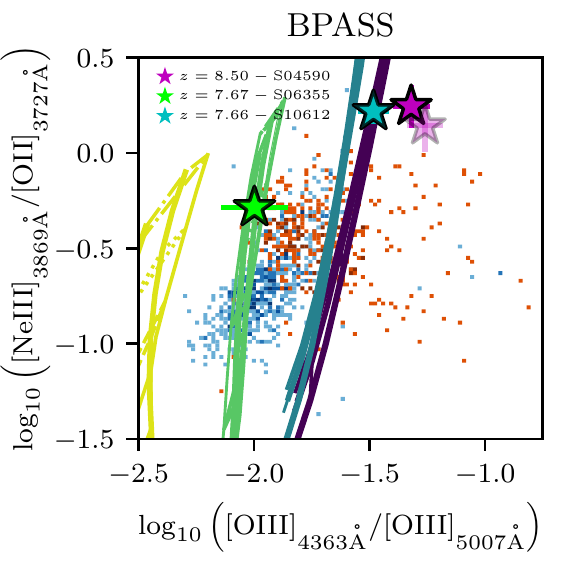}
\includegraphics[scale=1,trim={0 0.0cm 0cm 0cm},clip]{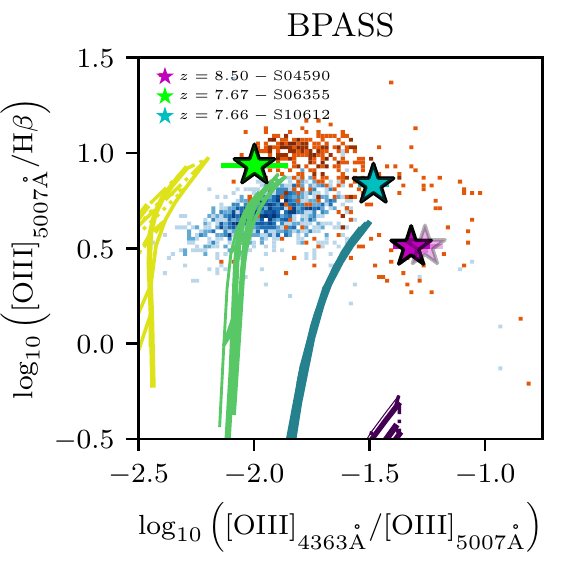}
}
\centerline{
\includegraphics[scale=1,trim={0 0.0cm 0cm 0cm},clip]{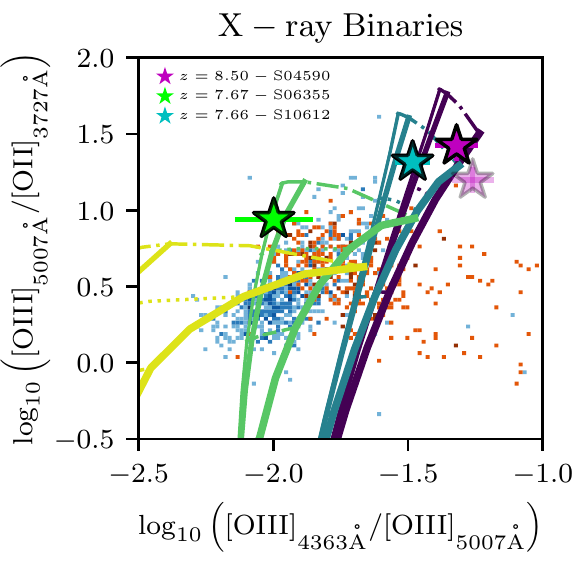}
\includegraphics[scale=1,trim={0 0.0cm 0cm 0cm},clip]{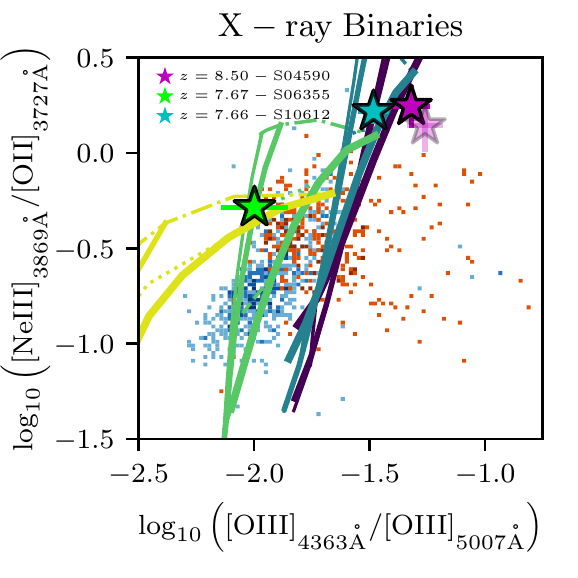}
\includegraphics[scale=1,trim={0 0.0cm 0cm 0cm},clip]{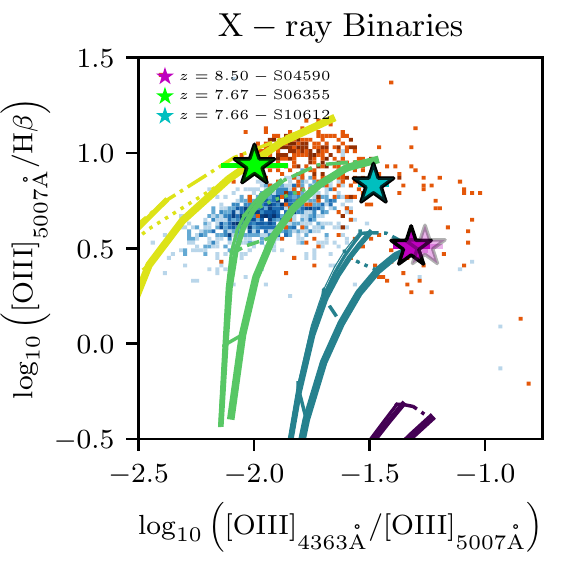}
}
\centerline{
\includegraphics[scale=1,trim={0 0.0cm 0cm 0cm},clip]{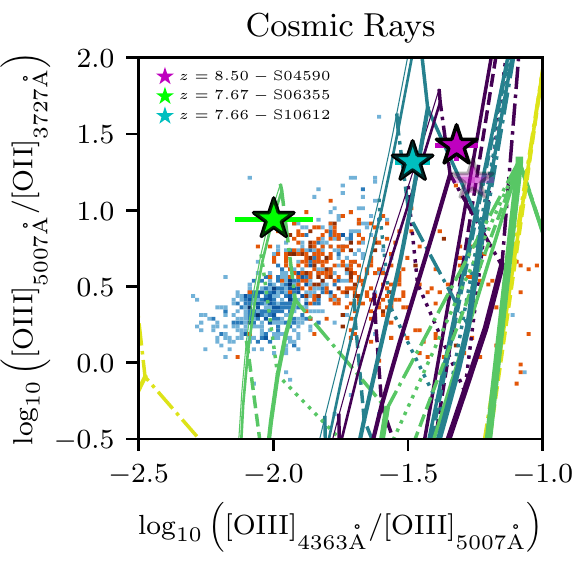}
\includegraphics[scale=1,trim={0 0.0cm 0cm 0cm},clip]{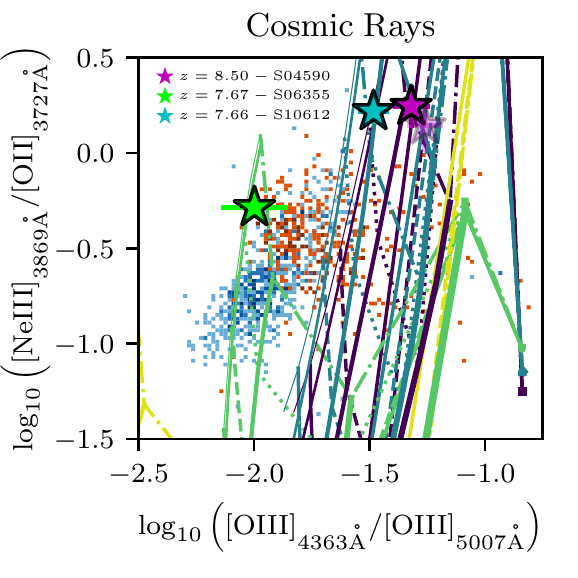}
\includegraphics[scale=1,trim={0 0.0cm 0cm 0cm},clip]{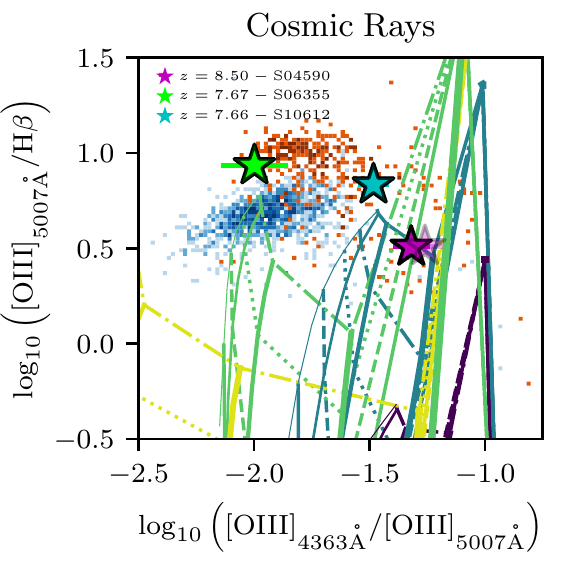}
}
\centerline{\includegraphics[]{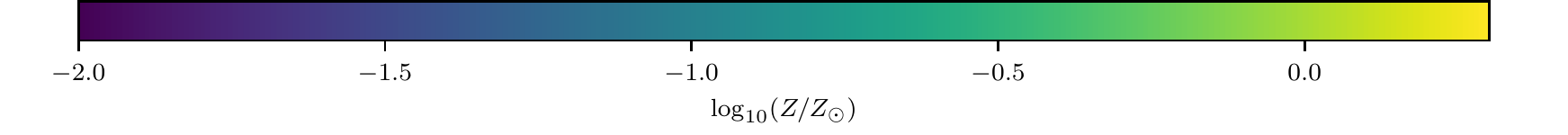}}
\caption{Various strong line diagnostics (Left: RO3 versus O32, Centre: RO3~versus [\neth]\,$\lambda 3869$/[\ot]\,$\lambda 3726, 3729$, Right: RO3 versus [\oth]\,$\lambda 5007$/H$\beta$) for the high redshift galaxies compared to {\small CLOUDY} models. The top row shows models that include only photoionization from stars, the middle row includes the impact of HMXBs, and the bottom row includes progressively stronger cosmic ray background intensities. The three high redshift galaxies are shown as coloured stars with error bars representing the 1$\sigma$ uncertainty on the flux ratios. Solid and translucent points represent intrinsic and dust-correct values, respectively for each galaxy. For reference, we show the location of SDSS galaxies and SDSS QSOs as the blue and orange background histograms. For all {\small CLOUDY} models, the colour of the line represents metallicity as given in the colour bar. Line style represents ionization parameter (solid, dashed, dotted, and dot-dashed correspond to $U$ values of $10^{-3.2}$, $10^{-2.6}$, $10^{-2.1}$, and $10^{-1.5}$, respectively, except for the cosmic ray panel where at a metallicity of $10^{-1}Z_{\odot}$ (blue), we have added an extra set of models with $U=10^{-0.5}$ shown as the solid line). Line thickness either represents stellar age (top), $L_X/{\rm SFR}$ (middle), or cosmic ray background intensity (bottom). In all panels, only a subset of {\small CLOUDY} models are shown for visual clarity.}
\label{cr_xrb}
\end{figure*}

\subsection{X-ray binaries}
High-energy photons from X-ray sources (including high-mass X-ray binaries or HMXBs) can potentially provide additional heating in the ISM \citep[e.g.][]{Lebouteiller2017,Lehmer2022}, above what one would expect from photoionization due to stars alone. Observational data has shown that the X-ray luminosity per unit star formation rate increases significantly at low metallicities \citep[e.g.][]{Mapelli2010,Basu-Zych2013, Saxena2021}. Furthermore, HMXBs have been postulated as an important ingredient to explain He{\small II} emission in low-metallicity star-forming galaxies \citep[e.g.][]{Schaerer2019,Senchyna2020,Saxena2020b,Umeda2022}. Since faint galaxies at $z>6$ are predicted to have low metallicities, HMXBs could be an important source of heating that is often missing from typical H{\small II} region photoionization or shock models. S04590 is predicted to have a much lower metallicity ($12+\text{log}({\rm O/H})=6.99\pm0.11$) than S06355 and S10612 ($12+\text{log}({\rm O/H})=8.24\pm0.07$, and $7.73\pm0.12$, respectively) \citep{Curti2022}, which means that HMXBs are likely to have a larger impact on S04590 compared to the other two galaxies.

To assess the impact of HMXBs, we have expanded our {\small CLOUDY} model grid, varying the ratio of X-ray luminosity\footnote{$L_X$ is defined as the X-ray luminosity in the range 2-10~keV.} to star formation rate ($L_X/{\rm SFR}$) in the range $10^{38}-10^{42}\ {\rm erg\ s^{-1}/M_{\odot}\ yr^{-1}}$, where observational constraints out to $z\sim5$ currently exist \citep{Saxena2021}. Details of these models are presented in Appendix~\ref{cloudy}. In the middle row of Figure~\ref{cr_xrb} we show three strong-line diagnostics along with a subset\footnote{We show a subset rather than the full grid for visual clarity.} of the results from our {\small CLOUDY} models. 

The models with weak X-ray luminosities are nearly identical (on the diagnostics that we consider) to the runs with only stars. Once the X-ray luminosity is increased, the diagnostics fan away from the photoionisation models that only include stars to high RO3 due to the additional heating from X-rays. As before, the models with lower metallicity tend to result in higher RO3 values due to the fact that cooling becomes less efficient with decreasing metallicity. 
Across all three diagnostics, the {\small CLOUDY} models with HMXBs can provide a reasonable explanation for the observed emission line ratios if high $L_X/{\rm SFR}>10^{41}\ {\rm erg\ s^{-1}/M_{\odot}\ yr^{-1}}$ is assumed. There remains some discrepancy in [\neth]~$\lambda$3869/[\ot]~$\lambda\lambda$3726, 3729, but we highlight how all three diagnostics and in particular, [\oth]~$\lambda$5007/H$\beta$, are significantly better at reproducing the observations compared to models with stars only. In summary, the inclusion of HMXBs with high $L_X/{\rm SFR}$ as predicted by extrapolating observed trends found $z\sim3-5$ \citep{Saxena2021} combined with low metallicity can plausibly reproduce the emission line ratios seen in S04590. Further tuning of the photoionization models (e.g. higher ionization parameters, changing the mass of the HMXB, non-solar abundance ratios) may improve the agreement. However, although physically motivated, we emphasize that this solution may not be unique because there are other heating mechanisms that can increase ISM temperature.

\subsection{High cosmic ray densities}
\cite{Ferland1984} considered cosmic rays as an additional heating mechanism to explain the observed RO3 ratios in low-redshift AGN. Assuming a simple model where galaxy volume scales with the size of the Universe (as $(1+z)^{-3}$)\footnote{Multiple observational studies have shown that galaxy size scales strongly with redshift \citep[e.g.][]{Shibuya2015,Allen2017}.}, it is plausible that the cosmic ray densities for high-redshift galaxies with similar star formation rates as the Milky Way may be $\sim1,000$ times higher than what is inferred from the galactic neighbourhood. The cosmic ray density should further scale with star formation rate if the cosmic rays are sourced by supernova explosions. \cite{Papadopoulos2011} showed that in cosmic ray-dominated regions where the cosmic ray density is $10^3-10^4$ times higher than the local neighbourhood, temperatures can easily rise to $5-10\times10^4$~K, which could help explain the high temperature needed to reproduce the observed RO3 in S04590. Such scenarios are predicted to operate in local ultraluminous infrared galaxies or high-redshift submillimeter galaxies. With the additional pressure and temperature, the stellar initial mass function is likely boosted in cosmic ray-dominated regions due to the increased Jeans mass, which aligns with certain predictions of a more top-heavy initial mass function at high-redshift \citep[e.g.][]{Katz2022b}.

To assess the impact of cosmic rays, we have further extended our {\small CLOUDY} grid by systematically increasing the strength of the cosmic ray background. The bottom row of Figure~\ref{cr_xrb} shows the results of these {\small CLOUDY} models compared to the three $z>7.5$ JWST galaxies. Here we can see fundamentally different behaviour compared to the other two grids. As the cosmic ray background is increased, RO3 increases due to the additional heating from the cosmic rays. However, initially O32, [\neth]~$\lambda$3869/[\ot]~$\lambda\lambda$3726, 3729, and [\oth]/H$\beta$ drop as well due to the increased temperature in the outer regions of the gas cloud allowing for more [\ot]~$\lambda\lambda$3726, 3729 and H$\beta$ emission. The models with $10^{-2}Z_{\odot}$, $U=10^{-1.5}$, and a cosmic ray background intensity that is $100\times$ greater than that of the Milky Way can plausibly explain both the O32, RO3, and [\neth]~$\lambda$3869/[\ot]~$\lambda\lambda$3726, 3729, but in this model, such low metallicity combined with the additional H$\beta$ emission means that [\oth]/H$\beta$ is systematically underpredicted. 

Interestingly, we can reproduce the behaviour of S04590 using cosmic rays in two ways. For each choice of ionization parameter and metallicity, there is a particular cosmic ray background intensity where the cloud switches from being ionization bounded to being density bounded. In this regime, O32 increases \citep[see e.g.][]{Nakajima2014} as the \ot~region in the nebula is destroyed. For the same reason, [\neth]~$\lambda$3869/[\ot]~$\lambda\lambda$3726, 3729 also increases. Finally, at sufficiently hot temperatures, [\oth]/H$\beta$ also increases because the recombination emissivity for H$\beta$ drops with temperature. If the cosmic ray background is very fine-tuned, the models that are right at the edge of the density and ionization bounded regimes can conspire to reproduce all of the flux ratios shown in Figure~\ref{cr_xrb}. Due to the fine tuning required, this is not our preferred solution, although we highlight that in this unique regime, cosmic rays will be the dominant driver of Lyman continuum leakage. To our knowledge, a cosmic ray-dominated ISM has not been considered as a mechanism for causing ionizing photons to leak from galaxies. 

More generally, if we increase the ionization parameter to $U=10^{-0.5}$, we can fit all the emission line ratios of S04590 shown in Figure~\ref{cr_xrb}. This solution still requires a cosmic ray background intensity that is more than $100\times$ that of the Milky Way; however, such a strong background is not unrealistic due to the small sizes and bursty nature of high-redshift galaxies as predicted by numerical simulations \citep[e.g.][]{Rosdahl2018}.

\section{Conclusions}
\label{sec:conclusions}
We have presented an analysis on the spectra of the first three $z>7.5$ galaxies observed with the \emph{JWST}/NIRSpec Micro-Shutter Assembly in the SMACS0723 field, realased as part of the Early Release Observations public data release.

We have compared emission line ratios from these high-redshift galaxies to various low-redshift galaxy (SDSS), QSO (SDSS), and potential high-redshift analogue \citep{Yang2017a,Yang2017b,Izotov2019,Amorin2015} populations. We find that the properties of the high-redshift galaxies are diverse and not a single analogue population well represents the flux ratios of all three systems across various diagnostics. Nevertheless, studying only oxygen and Balmer lines, the two galaxies (S06355 and S10612) at $z\sim7.6$ are consistent with the low-redshift analogues, while S04590 at $z=8.5$ seems to reside at the very edge of the low-redshift analogue distribution. 

S04590 is characterized by an extremely high [\oth]~$\lambda$4363/[\oth]~$\lambda$5007 ratio, which is at the upper limit of what is seen in very low metallicity galaxies in the local Universe \citep{Izotov2019}. We consider the possibility that additional physics beyond low-metallicity star formation is required to explain the nature of this source.

We show that neither AGN nor shock models can simultaneously reproduce the flux ratios observed in S04590. However, if we assume that high-mass X-ray binaries or cosmic rays contribute significantly to the heating of the ISM, we can reasonably reproduce many of the properties of S04590. In such a scenario, either the $L_X/{\rm SFR}$ is very high and consistent with extrapolated predictions from $z=3-5$ \citep[e.g.][]{Saxena2021}, or the cosmic ray background intensity is orders of magnitude greater than what is observed in the Milky Way, in line with expectations of decreasing galaxy size at high-redshift or generally more concentrated star formation. 

Our explanations of the observed flux ratios using cosmic rays or HMXBs are unlikely to be unique. In principle, any mechanism that can enhance the ionizing photon count and/or drive additional heating could potentially enhance RO3.

Due to the early nature of these observations, we remain cautious about over-interpreting the spectra, particularly the high RO3. Nevertheless, based on the available evidence, this work favours additional heating in the ISM, beyond photoheating from stars and photoelectric heating from dust grains to explain the high RO3. Cosmic rays or HMXBs could provide viable explanations for the high RO3 measured in S04590, indicating that we may need to revise our models of the ISM in the early Universe and how photons leak galaxies and contribute to reionization. \emph{JWST}/NIRSpec observations of larger samples of $z\gtrsim8$ galaxies will quantify whether or not these additional heating mechanisms are important in the high-redshift ISM.

\section*{Acknowledgements}
We thank the referee for their detailed comments that improved the manuscript. AS acknowledges financial support from European Research Council (ERC) Advanced Grant FP7/669253. AJC and AJB acknowledge funding from the ``FirstGalaxies'' Advanced Grant from the European Research Council (ERC) under the European Union's Horizon 2020 research and innovation programme (Grant agreement No. 789056). SA acknowledges funding from grant PID2021-127718NB-I00 by the Spanish Ministry of Science and Innovation/State Agency of Research (MCIN/AEI). RB acknowledges support from an STFC Ernest Rutherford Fellowship [grant number ST/T003596/1]. ECL acknowledges support of an STFC Webb Fellowship (ST/W001438/1). JW gratefully acknowledges support from the ERC Advanced Grant 695671, ``QUENCH'', and the Fondation MERAC.

\section*{Data Availability}
The 1D spectra used in this work are publicly available at \url{https://doi.org/10.5281/zenodo.6940561}. Any other data underlying this article will be shared on reasonable request to the corresponding author.

\bibliographystyle{mnras}
\bibliography{example} 

\appendix
\section{Description of CLOUDY Models}
\label{cloudy}
We have employed {\small CLOUDY} v17 \citep{Ferland2017} to calculate emission line fluxes for simplified photoionization models accounting for stellar radiation, X-ray binaries, and a cosmic ray background. For all models, we assume a 1~kpc thick shell with a constant density of 100~${\rm H\ cm^{-3}}$. In our fiducial set of models the inner radius of the slab is fixed to $10^{30}$~cm for a plane-parallel model and the ionization parameter is varied between $10^{-3.5}-10^{-1.5}$. The shape of the stellar SED is adopted from {\small BPASSv2.0} \citep{Stanway2016,Eldridge2016}. We assume an instantaneous burst and vary the age of the stellar population between $10^6-10^8$~yr. We vary the metallicity of the gas and stellar population between $10^{-2}Z_{\odot}$ and $2Z_{\odot}$. Metal abundances ratios are assumed to follow \citet{Asplund2009}. Dust grains are included using the {\small GRAINS ISM} command and the dust-to-gas-mass ratio is assumed to scale with metallicity. For all models, we account for metal depletion onto dust grains. Furthermore, we have turned off molecules.

To account for X-ray binaries, in addition to the stellar SED and CMB, we have added a modified blackbody spectrum \citep{Mitsida1984} that we assume captures the SED of XRBs. We fix the black hole mass to $25~{\rm M_{\odot}}$. Following \citet{Senchyna2020}, we assume the disk has a maximum radius of $10^4$~cm and compute the spectra using {\small ARES}, which is then combined with the stellar SED \citep{Mirocha2012}. Following observational constraints at high redshift \citep{Saxena2021}, we vary $L_X/{\rm SFR}$ in the range $10^{38}-10^{42}\ {\rm erg\ s^{-1}/M_{\odot}\ yr^{-1}}$, where $L_X$ is the X-ray luminosity in the range $2-10$~keV.

To assess the impact of cosmic rays, we have expanded our {\small CLOUDY} grid by removing the HMXB SED and replacing it with a cosmic ray background (using the {\small COSMIC RAYS BACKGROUND} command). We vary the cosmic ray background intensities from $1-10^{4.8}$ times the mean background strength of the Milky Way\footnote{We adopt a cosmic ray ionization rate of $2\times10^{-16}$ for hydrogen \citep{Indriolo2007}.}. For the cosmic ray runs, we also explore ionization parameters up to $10^{-0.5}$ at a metallicity of $10^{-1}Z_{\odot}$.

For all models, the calculation is stopped when the temperature reaches 4,000~K or when the outer radius of 1~kpc is reached. Note that these models are not designed to reproduce any particular galaxy population. Our goal is to qualitatively assess how HMXBs or cosmic rays may impact the relevant emission line ratios. 

\section{BPASS models with higher density and ionization parameter}
\label{bpe}
\begin{figure*}
\centerline{
\includegraphics[scale=1,trim={0 0.0cm 0cm 0cm},clip]{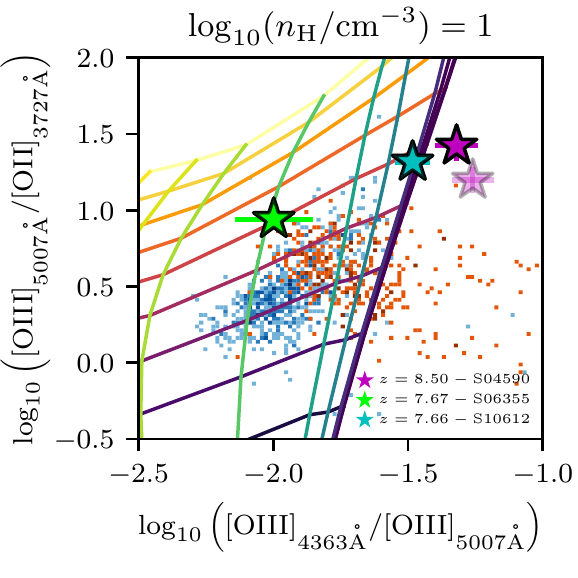}
\includegraphics[scale=1,trim={0 0.0cm 0cm 0cm},clip]{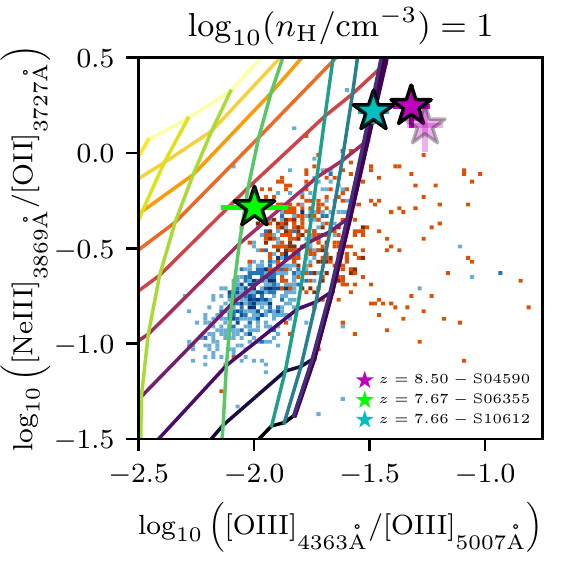}
\includegraphics[scale=1,trim={0 0.0cm 0cm 0cm},clip]{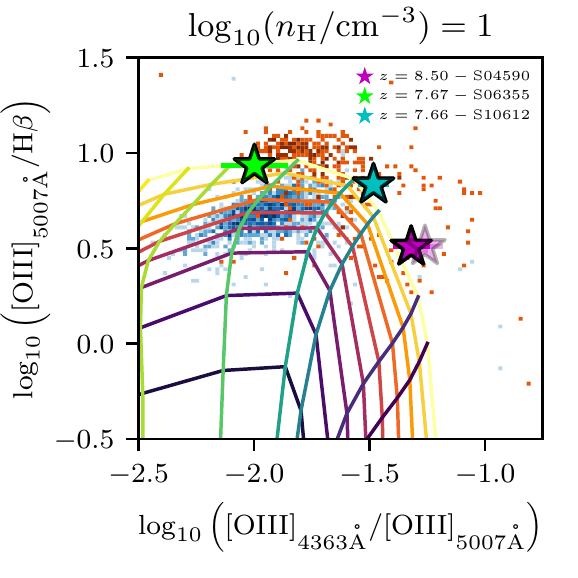}
}
\centerline{
\includegraphics[scale=1,trim={0 0.0cm 0cm 0cm},clip]{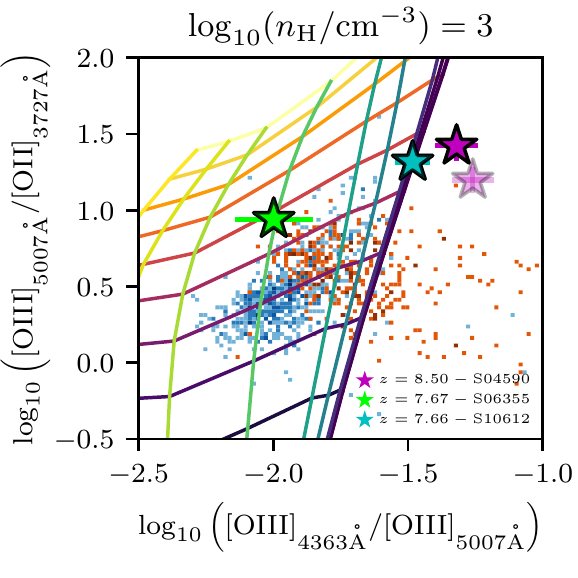}
\includegraphics[scale=1,trim={0 0.0cm 0cm 0cm},clip]{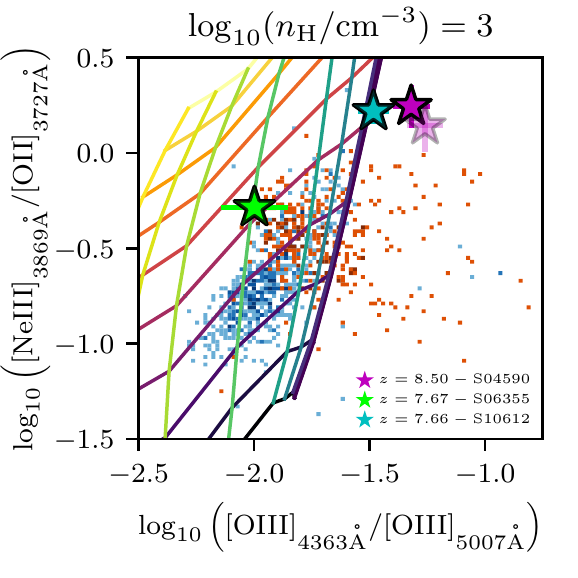}
\includegraphics[scale=1,trim={0 0.0cm 0cm 0cm},clip]{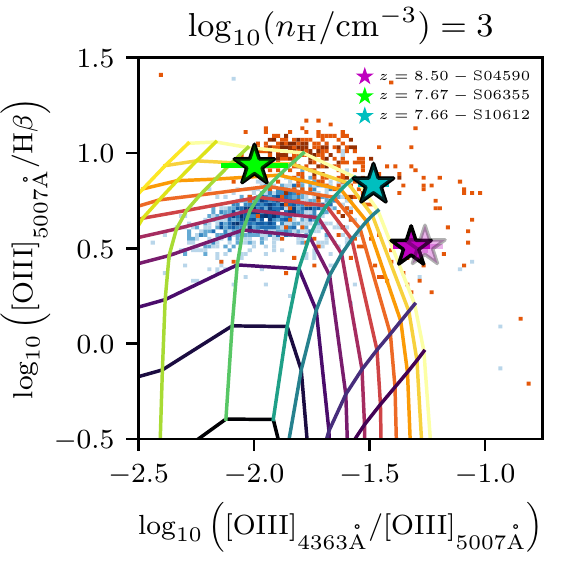}
}
\centerline{
\includegraphics[scale=1,trim={0 0.0cm 0cm 0cm},clip]{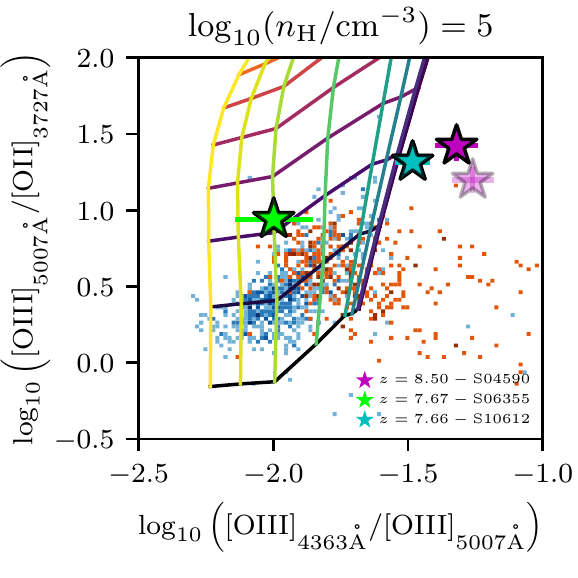}
\includegraphics[scale=1,trim={0 0.0cm 0cm 0cm},clip]{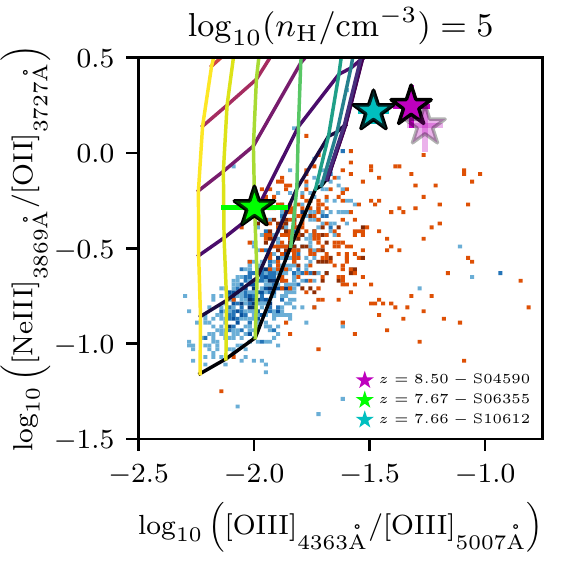}
\includegraphics[scale=1,trim={0 0.0cm 0cm 0cm},clip]{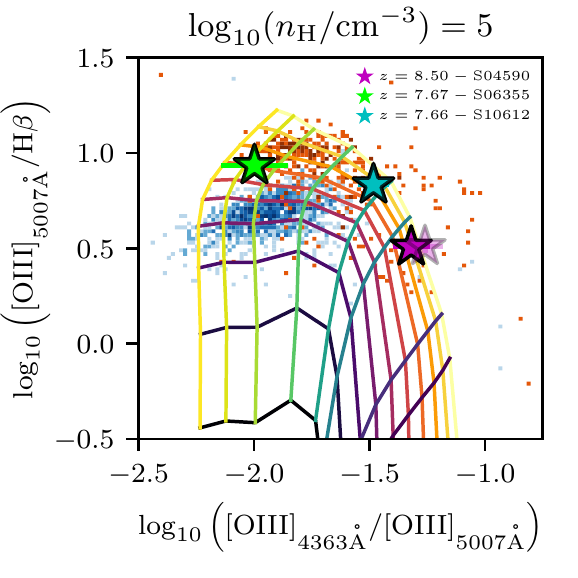}
}
\centerline{\includegraphics[]{./figures/metallicity_colorbar.pdf}}
\centerline{\includegraphics[]{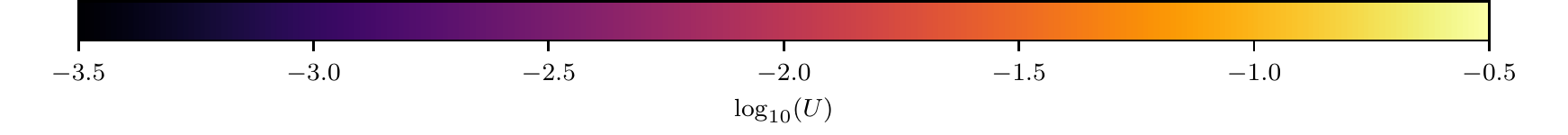}}
\caption{Various strong line diagnostics (Left: RO3 versus O32, Centre: RO3~versus [\neth]\,$\lambda 3869$/[\ot]\,$\lambda 3726, 3729$, Right: RO3 versus [\oth]\,$\lambda 5007$/H$\beta$) for the high redshift galaxies compared to {\small CLOUDY} models assuming a BPASSv2.0 SED a single burst with an age of 5~Myr. Each row shows runs with different gas densities while the different colours represent variations in ionization parameter or metallicity, as shown in the colour bars.}
\label{bpass_extra}
\end{figure*}

In Figure~\ref{bpass_extra}, we show {\small CLOUDY} model grids assuming a BPASSv2.0 SED a single burst with an age of 5~Myr extending up to densities of $10^5\ {\rm cm^{-3}}$ and ionization parameters up to $10^{-0.5}$. We show that for any combination of these parameters, the emission line properties of S04590 are not reproduced.

\section{Comparison with Shock Models}
\label{shock}
\begin{figure*}
\centerline{
\includegraphics[scale=1,trim={0 0.0cm 0cm 0cm},clip]{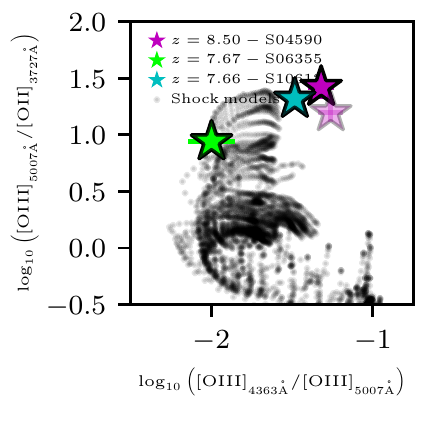}
\includegraphics[scale=1,trim={0 0.0cm 0cm 0cm},clip]{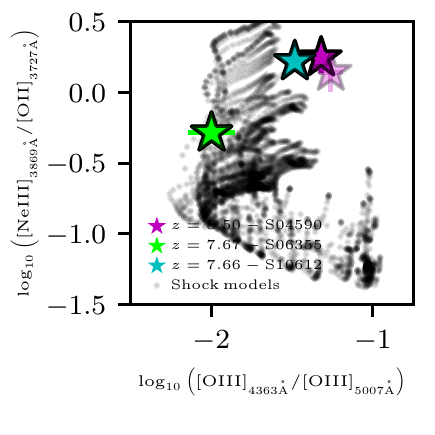}
\includegraphics[scale=1,trim={0 0.0cm 0cm 0cm},clip]{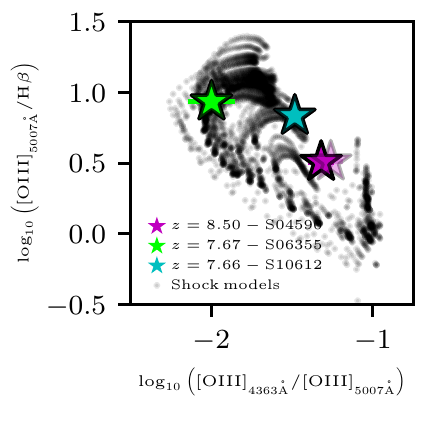}
\includegraphics[scale=1,trim={0 0.0cm 0cm 0cm},clip]{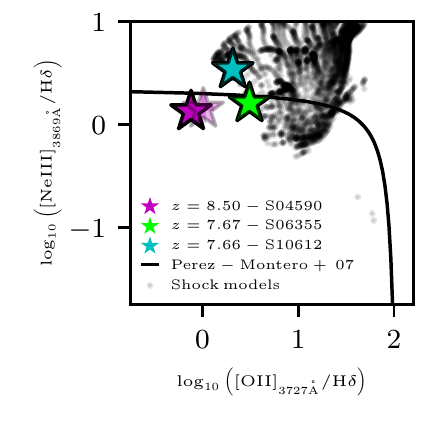}
}
\caption{(Far-left) RO3 versus O32. The three high redshift galaxies are shown as coloured stars with error bars representing the 1$\sigma$ uncertainty on the flux ratios. For comparison, we show MAPPINGS models of shocks from \protect\cite{Allen2008}. (Centre-left) RO3 versus [\neth]~$\lambda$3869/[\ot]~$\lambda\lambda$3726, 3729. (Centre-right) RO3 versus [\oth]~$\lambda$5007/H$\beta$. (Far-right) [\oth]~$\lambda\lambda$3726,3729/H$\delta$ versus [\neth]~$\lambda$3869/H$\delta$. The thick black line is the AGN/star-forming galaxy boundary as described in \protect\cite{PZ2007}.}
\label{shocks}
\end{figure*}

In Figure~\ref{shocks}, we compare the three $z>7.5$ galaxies with the shock models from \cite{Allen2008} across the four primary strong line diagnostics analyzed in this work.

\bsp	
\label{lastpage}
\end{document}